\documentclass[aps,prd,twocolumn,superscriptaddress,preprintnumbers,floatfix,nofootinbib]{revtex4-2}
\usepackage{showyourwork}
\usepackage{aas_macros}
\usepackage{graphicx}
\usepackage{amsmath}
\usepackage{siunitx}
\usepackage{placeins}
\usepackage{color}
\usepackage{standalone}
\usepackage{dcolumn}
\usepackage{tensor}
\usepackage{bm}
\usepackage{microtype}
\usepackage{etoolbox}
\usepackage{amssymb}
\usepackage{mathrsfs}
\usepackage{accents}
\usepackage[normalem]{ulem}
\usepackage[inline]{enumitem}
\usepackage[utf8]{inputenc}
\usepackage{lipsum}
\usepackage{booktabs}
\usepackage{wasysym}

\begin{document}

\title{Comment on ``Analysis of Ringdown Overtones in GW150914''}

\textbf{Comment on ``Analysis of Ringdown Overtones in GW150914''}
\emph{M.~Isi, W.~M.~Farr (Flatiron Institute)}

Reference \cite{Cotesta:2022pci} reanalyzes the GW150914 ringdown, arguing against the presence of an overtone and suggesting claims of its detection in \cite{Isi:2019aib} were driven by noise.
Here we point out a number of technical errors in that analysis, including a software bug, and show that features
highlighted as problematic are in fact expected and encountered in simulated data.
After fixes, the code in \cite{Cotesta:2022pci} produces results consistent with the presence of the overtone as in \cite{Isi:2019aib,Isi:2022mhy,Finch:2022ynt}, undermining several key claims in \cite{Cotesta:2022pci}.

Out of the different indicators for the overtone presented in
\cite{Isi:2019aib}, Ref.~\cite{Cotesta:2022pci} focuses on the posterior on the
overtone amplitude, $A_1$, as inferred by a two-tone model. For the
same model and data, the posterior is a mathematical object over which there
should be no disagreement. Yet, the $A_1$ distribution in \cite{Isi:2019aib}
visibly differs from the corresponding one in \cite{Cotesta:2022pci} (labeled $\Delta
t_{\rm start}^{\rm H1} = -0.72M$; $t_{\rm H1} = 1126259462.423 \, \mathrm{s}$), in
spite of \cite{Cotesta:2022pci} claiming to emulate the analysis in
\cite{Isi:2019aib}.

In a response \cite{Isi:2022mhy}, we subsequently failed to replicate the observations in \cite{Cotesta:2022pci} with a setup mimicking \cite{Isi:2019aib}; moreover, even when using the exact settings and software as in \cite{Cotesta:2022pci}, we remained unable to reproduce the posterior shown in \cite{Cotesta:2022pci} for the reference time in \cite{Isi:2019aib}.

\begin{figure}
  \includegraphics[width=\columnwidth]{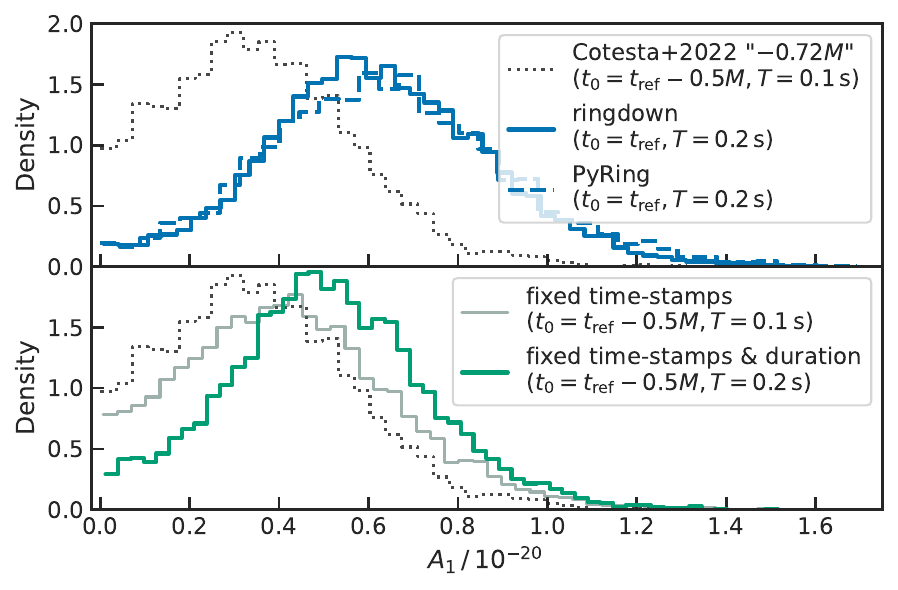}
  \caption{\emph{Top:} correct 16 kHz $A_1$ posterior at the reference time $t_{\rm ref}$ in \cite{Isi:2019aib} computed by \textsc{ringdown} (solid) and \textsc{pyRing} (dashed), contrasted with the incorrect posterior in \cite{Cotesta:2022pci} (dotted), as in the second panel of Supplement Fig.~4.
  \emph{Bottom:}
  besides failing to target $t_\mathrm{ref}$, results in~\cite{Cotesta:2022pci} suffered from a bug that mislabeled LIGO timestamps and analyzed insufficient data (length $T = 0.1\,{\rm s}$); the bottom panel shows the effect of fixing each of these issues, to eventually yield the true $A_1$ posterior at $t_0 = t_{\rm ref} - 0.5 M$ (thick green), which differs only slightly from the intended target at $t_0 = t_{\rm ref}$ (top blue).
  \script{figure1.py}
  \label{fig:reftime}
}
\end{figure}

We have now established that the analysis in \cite{Cotesta:2022pci} that is reported as
analogous to that in \cite{Isi:2019aib} actually
(i) started at a different time ($\Delta t_{\rm start}^{\rm H1} = -1.2 M$; $t_{\rm H1} = 1126259462.422828 \, \mathrm{s}$),
(ii) used too short a segment of data given the response of the whitening filter, and
(iii) was affected by a coding bug that attributed erroneous timestamps to the LIGO data \cite{bug}.
Correcting all this, the codes used in
\cite{Isi:2022mhy} and \cite{Cotesta:2022pci} (\textsc{ringdown} \cite{ringdown}
and \textsc{pyRing} \cite{pyRing_soft}) yield indistinguishable
results, favoring the presence of the overtone at the time in \cite{Isi:2019aib}
(Fig.~\ref{fig:reftime}). This agrees with the independent amplitude
posteriors in \cite{Finch:2022ynt}, and contradicts the claim in
\cite{Cotesta:2022pci} that ``around the peak, the Bayes factor does not
indicate the presence of an overtone.''

\begin{figure}
  \includegraphics[width=\columnwidth]{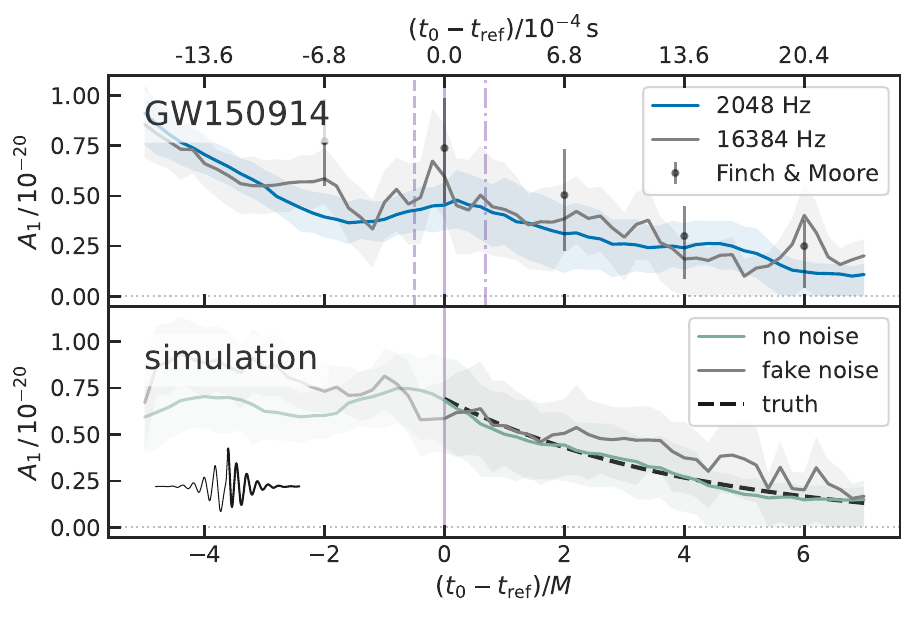}
  \caption{$A_1$ posterior median (line) and 68\% HPD (band) for different fit times $t_0$ relative to $t_{\rm ref} = 1126259462.423 \, \mathrm{s}$ (abscissa). \emph{Top}: GW150914 analyzed at 2 kHz (blue) and 16 kHz (gray), compared to the result in \cite{Finch:2022ynt} (dots); a dashed line marks the time used in \cite{Cotesta:2022pci} and incorrectly labeled as $t_{\rm ref}$ (solid line); a dot-dashed line shows the the peak time estimate in \cite{Cotesta:2022pci}.
  \emph{Bottom}: simulation of two-damped sinusoids drawn from a GW150914 posterior without noise (green) and in synthetic LIGO noise (gray) at 16 kHz; to avoid an abrupt start, the injection is a ``ring-up'' before $t_0 = t_{\rm ref}$, as in Fig.~10 in \cite{Isi:2021iql}.
  The GW150914 analyses prefer the overtone over a broad set of times around the inferred peak (top), and the posterior tracks the decay of the mode plus, for the 16 kHz runs, additional jitter from high frequency noise, consistent with damped sinusoids in Gaussian noise (bottom).}
  \script{figure2.py}
  \label{fig:variation}
\end{figure}

Both \textsc{pyRing} and \textsc{ringdown} favor $A_1 > 0$ over a broad range of start times after the peak, as identified in either  \cite{Isi:2022mhy} or  \cite{Cotesta:2022pci}:
the significance, as evaluated from the $A_1$ posterior, does not ultimately drop below $2\sigma$ until {${\sim}10\,\mathrm{ms}$} after the reference time in \cite{Isi:2022mhy}, consistent with the expected decay of this mode (Fig.~\ref{fig:variation}, top).

Reference \cite{Cotesta:2022pci} argues that the overtone detection in \cite{Isi:2019aib} was ``noise dominated'' because the ``amplitude is sensitive to changes in the starting time much smaller than the overtone damping time.''
However, the variation that remains after fixing the timing problems in \cite{Cotesta:2022pci} is expected: because different data are processed for different start times, the posterior will jitter due to high frequency noise.
With data sampled at 16 kHz, as in \cite{Cotesta:2022pci}, the $A_1$ posterior may vary over times as short as $\approx 0.1\, \mathrm{ms}$---faster than for the 2 kHz sampling rate in \cite{Isi:2019aib,Isi:2022mhy}.
We demonstrate this with simulated damped sinusoids in synthetic noise, showing similar fluctuations as for GW150914 (Fig.~\ref{fig:variation}, bottom).

In summary, both key points in \cite{Cotesta:2022pci} (namely, that there is no evidence for $A_1 > 0$ at any time after the peak, and that posterior variability indicates the result was noise-dominated) were drawn from faulty posteriors affected by erroneous settings and a software bug.
These errors conspired to spuriously exacerbate the $A_1$ posterior width and sensitivity to start time.
On fixing these errors, the code used in \cite{Cotesta:2022pci} returns results in agreement with the posteriors in the rest of the literature and in tension with central claims in \cite{Cotesta:2022pci} itself.

\bibliography{bib}

\end{document}